\documentclass[twocolumn,prl,aps,floatfix,epsf,psfig,groupaddress,amssymb]{revtex4}
\usepackage{graphicx,subfigure}
\usepackage{color,soul}
\usepackage{epsfig}
\usepackage{mathtools}
\DeclarePairedDelimiter{\evdel}{\langle}{\rangle}
\newcommand{\ev}{\operatorname{}\evdel}
\usepackage{amsmath}
\usepackage{float}
\begin{document}
\title{Defect induced Anderson localization and magnetization in graphene quantum dots}

\author{A. Alt{\i}nta\c{s}}
\author{A. D. G\"u\c{c}l\"u}

\affiliation{Department of Physics, Izmir Institute of Technology, IZTECH,
  TR35430, Izmir, Turkey}

\date{\today}

\begin{abstract}
We theoretically investigate the effects of atomic defect related
short-range disorders and electron-electron interactions on Anderson
type localization and the magnetic properties of hexagonal armchair
graphene quantum dots using an extended mean-field Hubbard model and
wave packet dynamics for the calculation of localization lengths. We
observe that randomly distributed defects with concentrations between
1-5\% of the total number of atoms leads to localization alongside
magnetic puddle-like structures.  Although the localization lengths
are not affected by interactions, staggered magnetism and localization
are found to be enhanced if the defects are distributed unevenly
between the sublattices of the honeycomb lattice. 

\end{abstract}
\maketitle

\section{Introduction}


Graphene\cite{novoselov2004electric,novoselov2005two,zhang2005experimental,rycerz2007valley,Potemski+deHeer+06},
a promising single-layer material for electronics applications, has
been getting increasing interest in understanding and engineering its
properties at the nanoscale to form graphene nanoribbons and
dots. Indeed, electronic, magnetic and optical properties of graphene can be
tuned by changing edge, shape, doping and number of
layers\cite{guclu+book14,Trauzettel+07,Schnez+Ensslin+08,Wimmer+10,Ihn+Ensslin+10,mueller_yan_nanolett2010,Hamalainen+Liljeroth+11,Subramaniam+12,Olle+Gambardella+12,isil+14,Ezawa+07,FRP+07,Wang+Meng+08,AHM+08,Guclu+09,Potasz+10,Zarenia+11,ma+12,Guclu+Potasz+Hawrylak+2013,Szalowski+13,altintas+2017,modarresi+guclu17,ozdemir+16,guclu2016,sevincli+08,guclu+bulut15,nanostructures1,nanostructures2,nanostructures3,nanostructures4,
nanostructures5,nanoribbonedge5,YuanyuanSun+2017}. On
the other hand, introducing
adatoms\cite{AdatomExperimet1,AdatomExperimet2,AdatomExperimet3,AdatomExperimet4,AdatomExperimet5,AdatomExperimet6}
or vacancies\cite{Mao+2016,VacancyExperimet1,VacancyExperimet2,VacancyExperimet3,AdatomandvacancyExperimet1}
 can also significantly affect its physical properties. For
example, a dramatic increase in resistivity of graphene,
metal-to-insulator (localization) behavior and magnetic moment
induction which led to spin split state at the Fermi energy were
observed in several experimental works by introducing hydrogen adatoms
on graphene
\cite{AdatomExperimet3,AdatomExperimet5,MetaltoInsulatorExperiment1}. Additionally,
local magnetism due to vacancies created by irradiation of graphene
samples were
detected\cite{VacancyExperimet1,AdatomandvacancyExperimet1}.

There have been many theoretical attempts to explain induction of
metal-to-insulator transition and magnetism brought
about by adatom or vacancy related disorders in graphene
structures\cite{AdatomVacancyDFT1,VacancyMFH1,AdatomMFH1,VacancyMixedTheory1,AdatomDFT1,AdatomDFTMFH1,VacancyDFT1,AdatomDFTTB,AdatomDFTMFH2,AdatomDFTMFH3,AdatomAndersonLocaTB1,sevincli+08,guclu+bulut15}. For
instance, ferromagnetic or antiferromagnetic behavior of
quasilocalized states can be induced by introducing two atomic defects
on the same or opposite sublattices of the honeycomb
lattice. Furthermore, it was found that vacancy related sublattice
imbalance which leads to total spin $S\neq0$ can induce global
magnetism predicted by Lieb and sublattice balance which leads to
total spin $S=0$ can induce local magnetism by using mean-field
Hubbard model for graphene ribbons\cite{VacancyMFH1,LiebsTheorem}. On
the other hand, Schubert \textit{et al.}\cite{AdatomAndersonLocaTB1}
used a tight-binding (TB) model ignoring magnetic effects to show that
low concentrations of randomly distributed hydrogen adatoms lead
to metal-to-insulator transition in graphene, although alongside
formation of electron-hole puddles that tend to suppress Anderson
localization\cite{Anderson}.

An interesting and natural question to ask is whether the magnetic and
localization properties are affected by each other, which, to the best
of our knowledge, remains unaddressed presumably due to difficulties
in incorporating electron-electron interactions in large size
systems. In this work, in order to find out the role of atomic defects
in both the localization of electronic states and the magnetic
behavior at the nanoscale, we perform mean-field Hubbard (MFH)
calculations for medium sized graphene quantum dots (GQD). More
specifically, we focus on hexagonal shaped GQDs with armchair edges
which are, unlike zigzag edges, free of magnetized edge effects. Thus
hexagonal armchair GQDs allow for an unbiased investigation of defect
induced magnetization and provide a link between nanosize and bulk
limits. The localization properties are investigated using wave function
dynamics. We show that localization of electronic states can occur due
to atomic defects, together with formation of magnetic puddles. We
found that, although the localization lengths are not affected by
interactions for evenly distributed defects between the two
sublattices, an uneven distribution between the two sublattices can
significantly enhance both the localization and the magnetization.
Surprisingly, no spin dependent localization lengths were observed.

The structure of the paper is as follows. In Sec. II, we describe our
model Hamiltonian including electron-electron interaction and the
computational methods that we use in order to compute magnetic and
localization properties of hexagonal armchair GQDs. The computational
results are presented in Sec. III.  Finally, Section IV provides
summary and conclusion.

\begin{figure}[]
\includegraphics[scale=0.3]{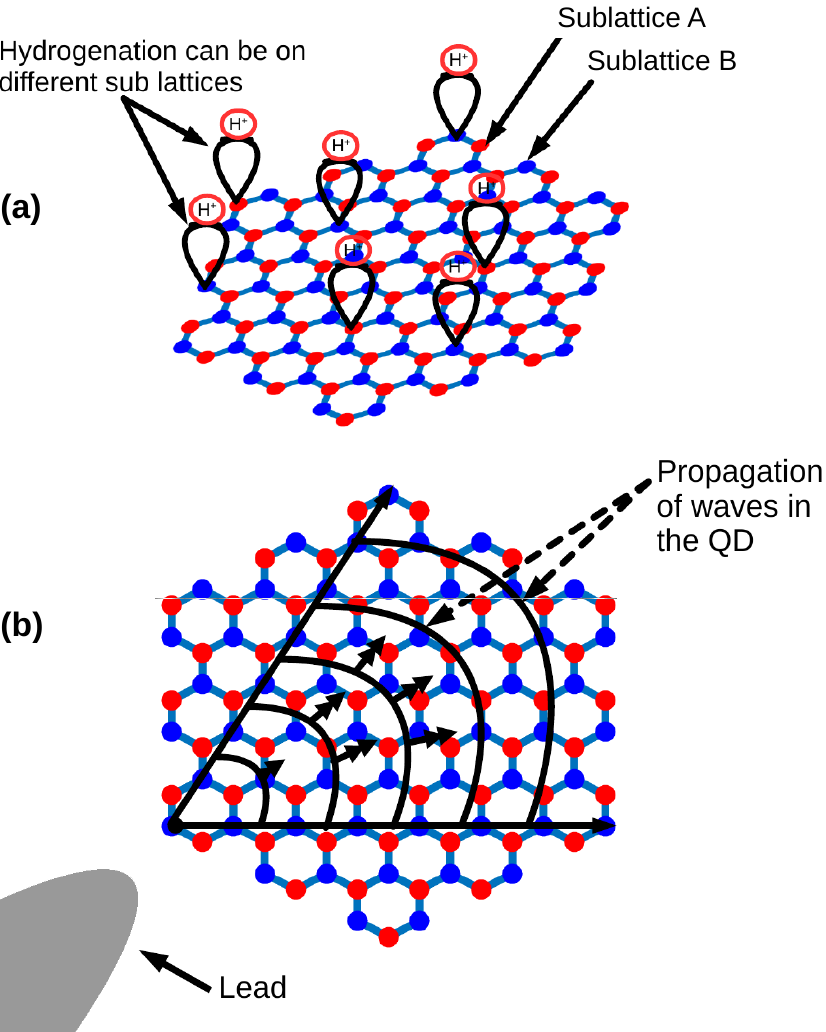}
\caption{(Color online) (a) Hydrogenation as a short-range disorder model
  on a hexagonal armchair edged GQD. (b) Cartoon of propagation of
  waves corresponding to an electron injected from one corner of the
  QD.  }
\label{fig1:result}
\end{figure}

\section{Method and Model}
We use the extended one-band MFH model where the single electron states 
can be written as a linear combination of $p_z$ orbitals on every
carbon atom since the sigma orbitals are considered to
be mainly responsible for mechanical stability of graphene. 
Within the extended MFH model, Hamiltonian can be written as:

\begin{align}
H_{MFH} =& \sum_{ij\sigma} ( t_{ij} c^{\dagger}_{i\sigma} c_{j\sigma} + h.c) \nonumber \\
&+ U\sum_{i\sigma} (\ev{n_{i\sigma}} - \frac{1}{2})n_{i\bar{\sigma}} + \sum_{ij\sigma} V_{ij} (\ev{n_{j}}-1)n_{i\sigma} 
 \nonumber \\ 
&
\end{align}
\noindent
where the first term represents the TB Hamiltonian and $t_{ij}$ are
the hopping parameters given by $t_{nn}=-2.8$ eV for nearest
neighbours and $t_{nnn}=-0.2$ eV for next
nearest-neighbours\cite{hopping1}. The $c^{\dagger}_{i\sigma}$ and
$c_{i\sigma}$ are creation and annihilation operators for an electron
at the $i$-th orbital with spin $\sigma$, respectively. Expectation
value of electron densities are represented by $\ev{n_{i\sigma}}$. The
second and third terms represent onsite and long range Coulomb
interactions, respectively.  We note however that, the inclusion of
long-range Coulomb interactions did not significantly affect the
numerical results in this work. This is in contrast with our previous
work\cite{altintas+2017} on the investigation of long range scatterers
which cause strong density modulations, leading to non-negligible
long-range Coulomb interactions. We take onsite interaction parameter
as $U=16.522/ \kappa$ eV and long-range interaction parameters
$V_{ij}=8.64/ \kappa$ and $V_{ij}=5.33/ \kappa$ for the first and
second nearest neighbours with effective dielectric constant
$\kappa=6$\cite{dielectricconstant1}, respectively. Distant neighbor
interaction is taken to be $1/d_{ij} \kappa$ and interaction matrix
elements are obtained from numerical calculations by using Slater
$\pi_z$ orbitals \cite{Slatermatrix1}. To account for short-range
disorder effects (which may be due to vacancies or hydrogen
adatoms. See Fig.~\ref{fig1:result}a), we simply remove corresponding
$p_z$ orbital sites. This model assumes that sp2 hybridization of
atoms neighboring the defect and the honeycomb lattice are not
distorted.

\begin{figure}[t]
\includegraphics[scale=0.23]{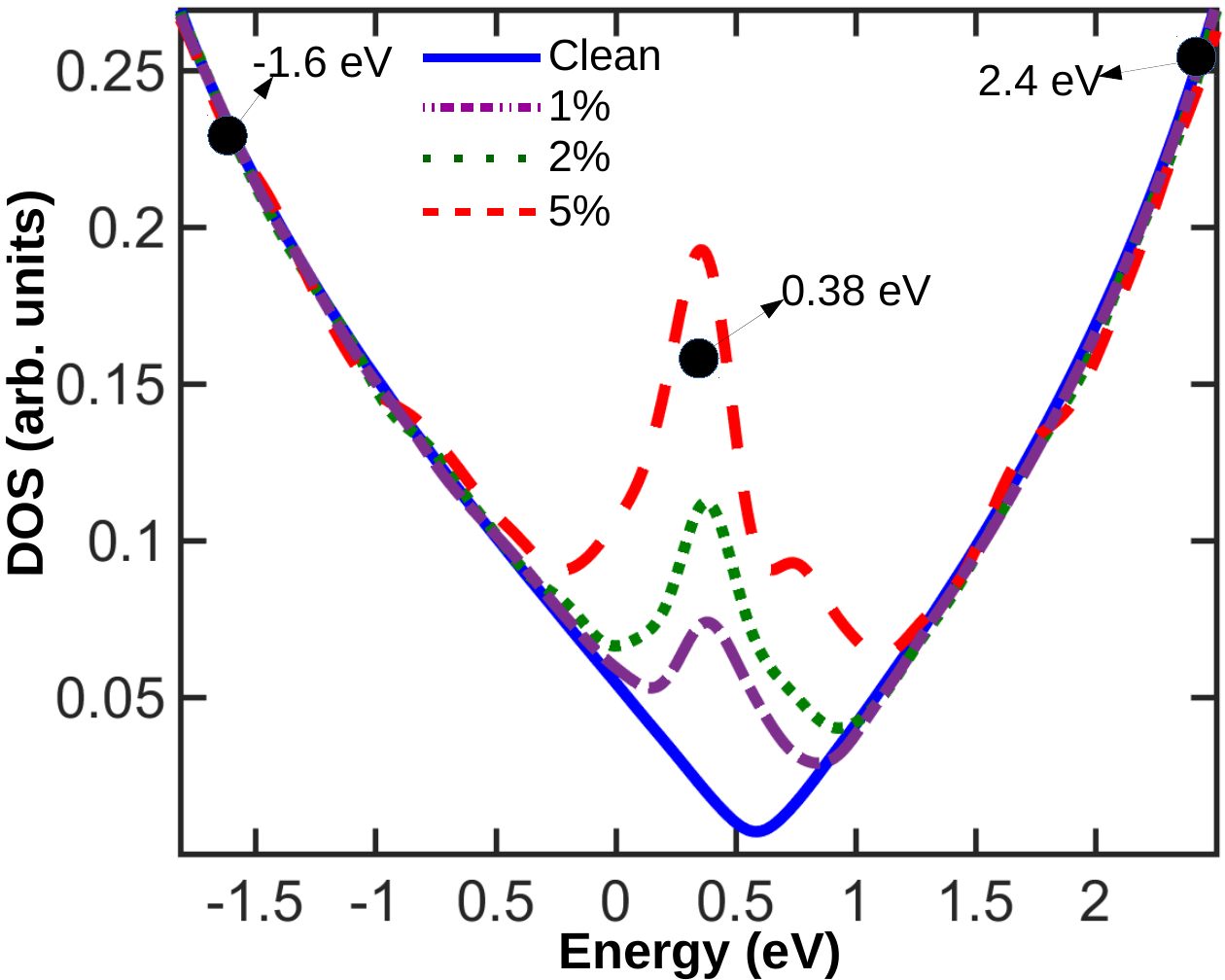}
\caption{(Color online) Density of states obtained by TB model for
  clean (solid blue line),\%1 (dotted and dashed purple line), \%2
  (dotted green line) and \%5 (dashed red line) disordered dot. Big
  black dots show incoming electrons with specific energy. One
  configuration is shown for each percent of disorder since other 19
  configurations show similar behavior.}
\label{fig2:result}
\end{figure}

\begin{figure}[b]
\includegraphics[scale=0.24]{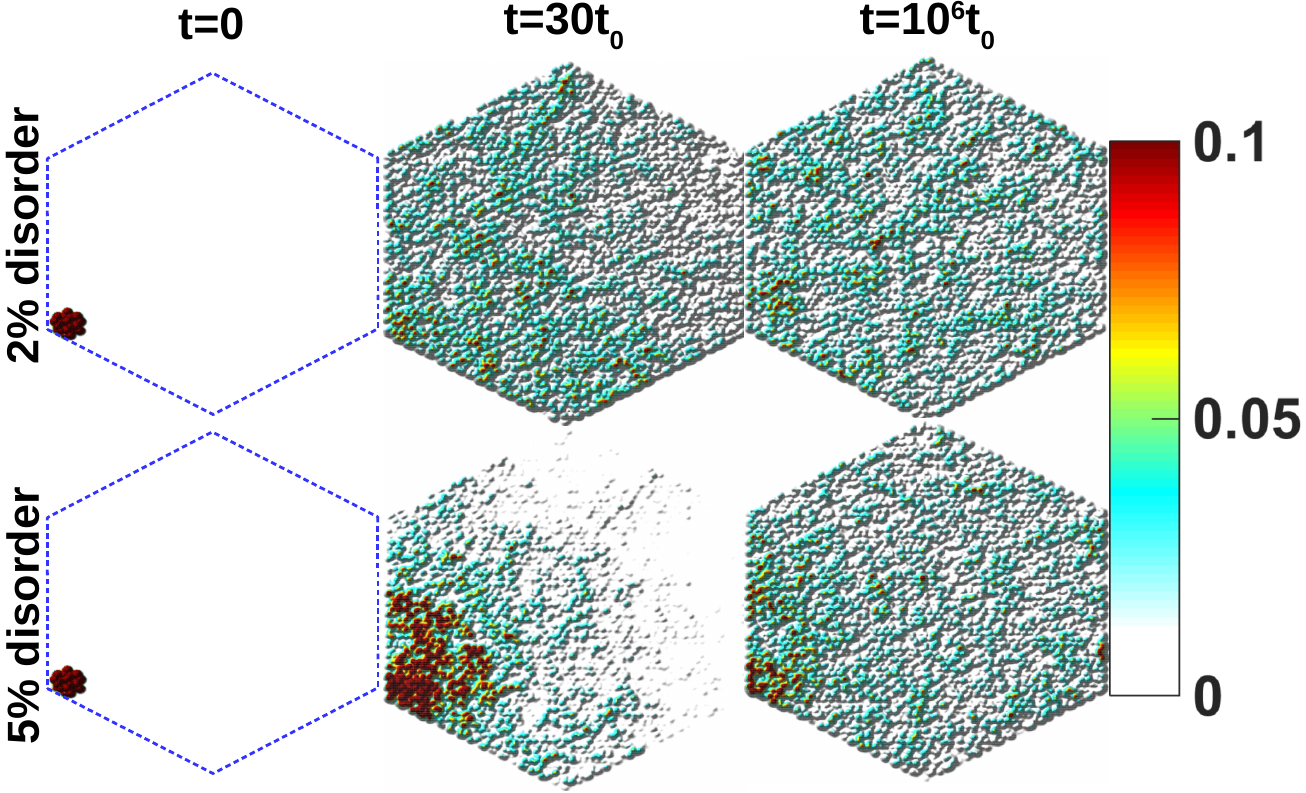}
\caption{(Color online) Time evolution of electronic density obtained by
  TB model for disordered GQD. Electron injected through one corner of the hexagonal QD has average energy of $\left\langle E_{i} \right\rangle= 0.38$ eV. Each column panels show the snapshot of
  wave packet propagation at, from left to right, time $t=0$, $t=30t_0$
  and $t=10^6t_0$. Top and bottom panels are for 2\% and 5\% disorder
  distributions, respectively.}
\label{fig3:result}
\end{figure}

\begin{figure*}[t]
\includegraphics[scale=0.45]{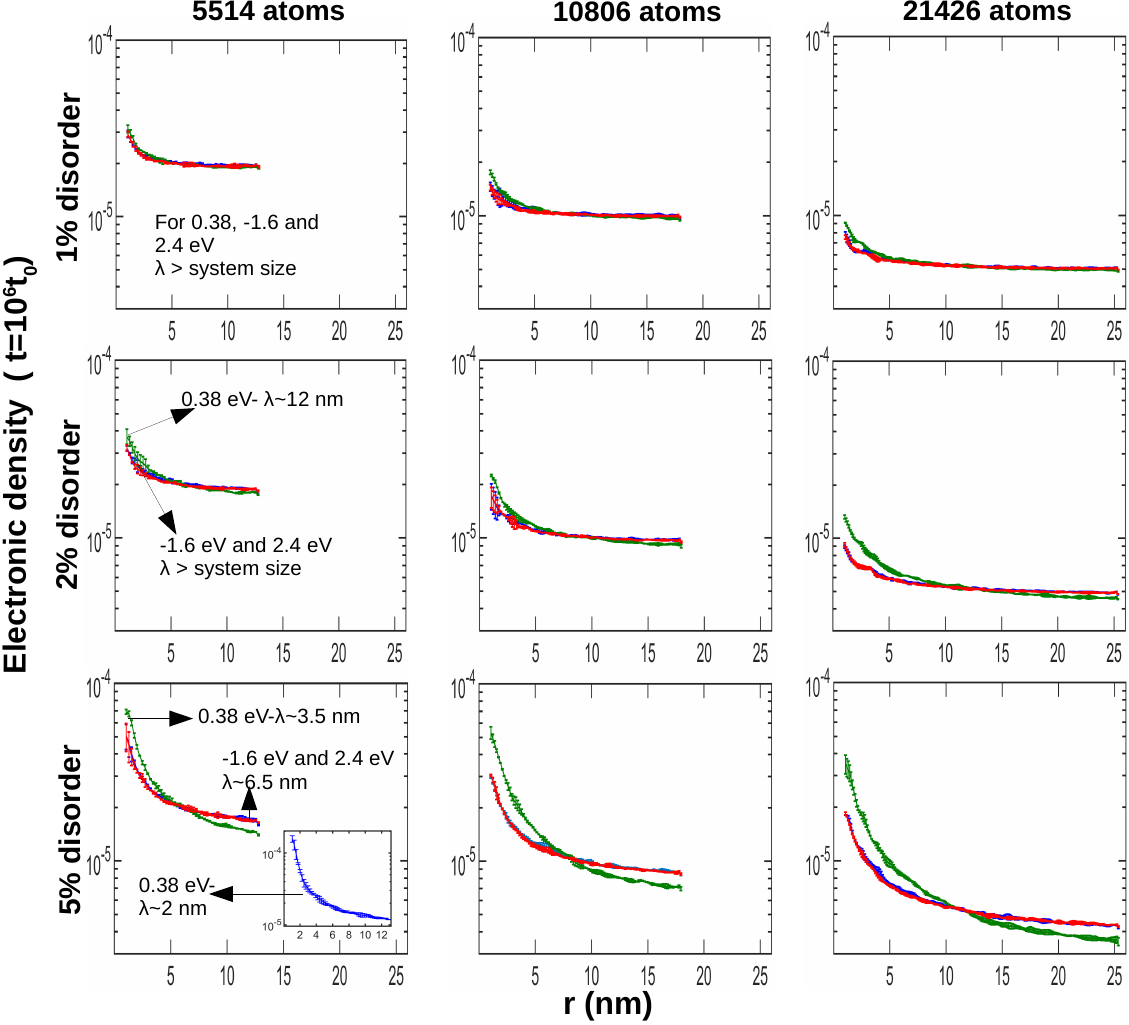}
\caption{(Color online) Time and angle averaged electronic densities
  for different incoming wave packet energies, obtained by TB
  method. Each column corresponds to different sizes of GQDs and row
  corresponds to 1\%, 2\% and 5\% percent of randomly created disorder
  (evenly distributed between sublattice A and B). The horizontal axis
  represents propagation of the wave along the length which starts
  from the contacted edge of the QD to the opposite edge (see
  Fig. 1b). Propagation time is taken long enough (order of nano
  second) so that wave can be considered as quasi-stationary. Each
  curve represents average of 20 different configurations with
  corresponding error bars. Localization lengths of GQD containing
  5514 atoms are only shown since localization lengths of larger QDs
  have similar values for the same rows. Inset shows uneven
  distribution of defects (100-0\%). At $1\%$
  defect concentration, size effects dominate the densities and
  localization length is larger than the system size even for the
  largest QD (25 nm wide) and the energy dependence is weak. As the
  defect concentration is increased to $2\%$, localization length
  $\lambda\sim 12$ nm for injected electron having energy $\left\langle E_{i} \right\rangle = 0.38$ eV for all QD
  sizes. For energies $\left\langle E_{i} \right\rangle =$ -1.6 and 2.4 eV, $\lambda$ exceeds the system
  size. Finally, increasing defect concentration to 5\% decreases
  localization length to $\lambda\sim 3.5$ nm for $\left\langle E_{i} \right\rangle = 0.38$ eV for all
  QD sizes. Additionally, we start to observe localization
  ($\lambda\sim 6.5$ nm) for the energies $\left\langle E_{i} \right\rangle =$ -1.6 and 2.4 eV.}
\label{fig4:result}
\end{figure*}

\begin{figure}[t]
\includegraphics[scale=0.25]{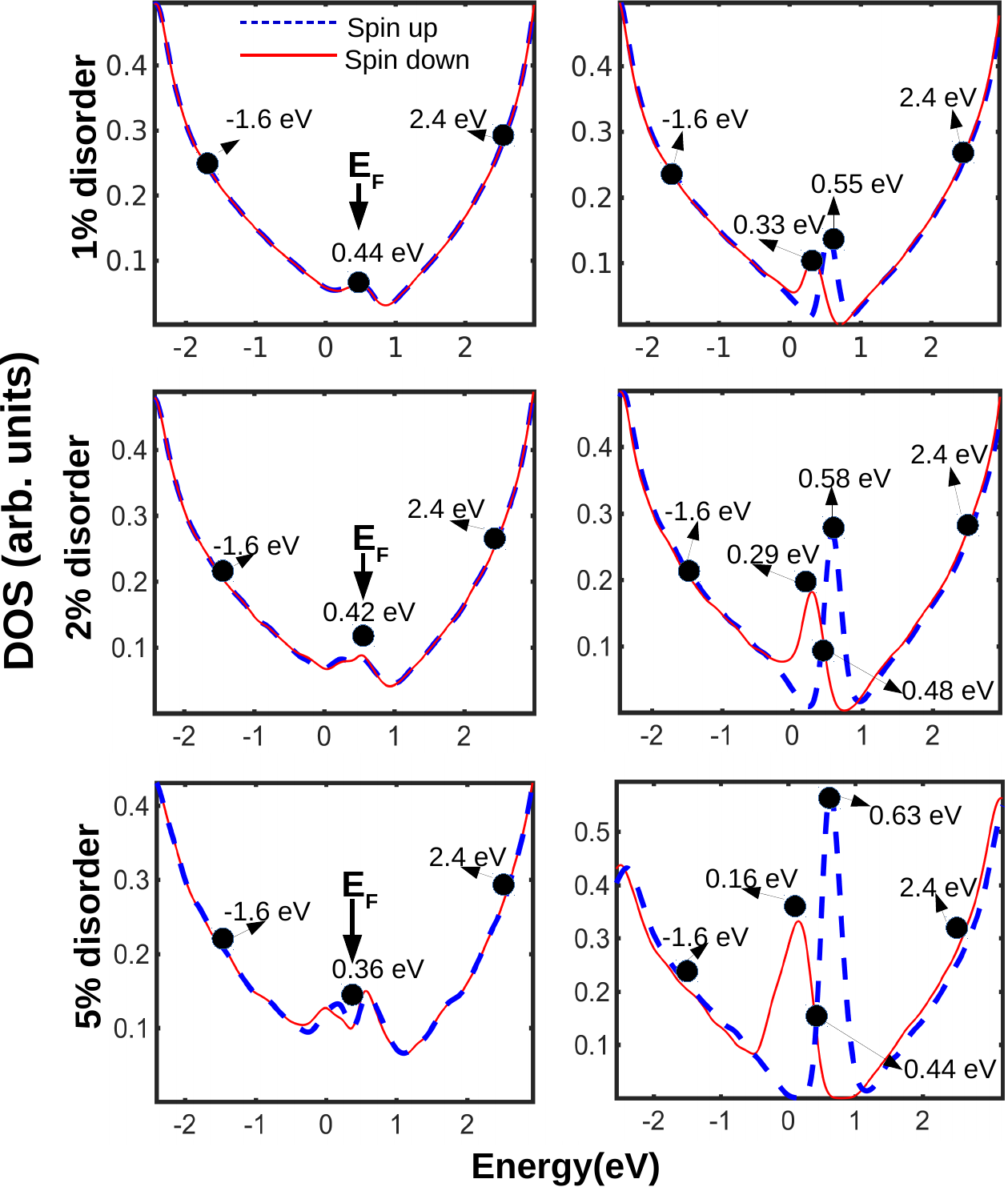}
\caption{(Color online) Density of states for spin down (red solid
  line) and spin up (dashed blue line), for 1\% (upper panels), 2\%
  (middle panels) and 5\% (lower panels) disorder concentrations,
  randomly distributed among each sublattice as 50\% (50\%) (left
  panels) and 100\% (0\%) (right panels) for sub lattice A (B).  Black
  dots show incoming electron's energy to be used in localization
  length calculations. $E_F$ indicates the Fermi energy. A clear spin splitting is observed in DOS for the right panels, a signature of ferromagnetic coupling.}
\label{fig5:result}
\end{figure}

\begin{figure*}[th]
\includegraphics[scale=0.45]{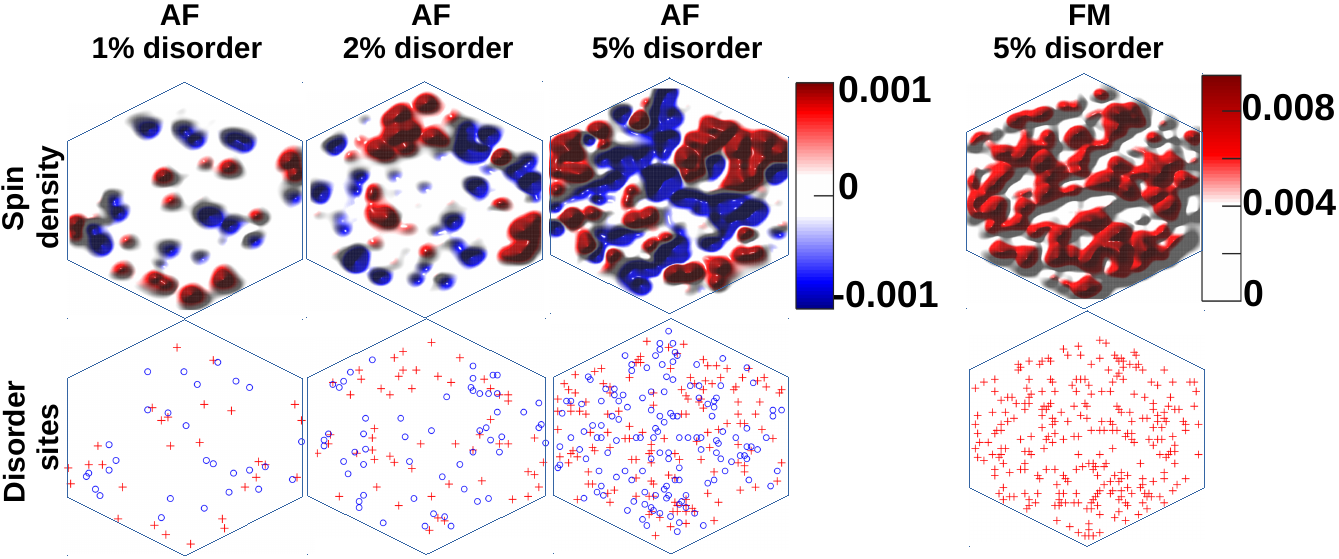}
\caption{(Color online) Magnetic puddle formation in antiferromagnetic
  (AF) and ferromagnetic (FM) GQDs. Disorders are randomly distributed
  among each sublattice as 50\% (50\%) (first three panels) and 100\%
  (0\%) (last panel) for sub lattice A (B) . Upper panels show spin
  density profile and red (blue) regions represent either spin up or
  down electrons. The corresponding disorder sites are shown in lower
  panels by blue circles and red crosses. When the system is
  antiferromagnetic (for even number of sublattice A and B defects),
  statistical distribution of defects gives rise to formation of
  magnetic puddles with opposite signs (shown in red and blue colors
  online). The system is ferromagnetic if the defects are distributed
  solely on on sublattice A.}
\label{fig6:result}
\end{figure*}

A critical step in the numerical calculations is the initial guess
state used for the self-consistent diagonalization of the MFH
Hamiltonian, as there is a high risk of getting stuck in a local
energy minimum for systems with several thousands of atoms. Local
version of Lieb's theorem provides a convenient way to generate the
initial state. According to Lieb's theorem\cite{LiebsTheorem}, if
there is an overall imbalance between the number of A and B sublattice
atoms in a bipartite lattice, a finite magnetic moment $(N_A-N_B)/2$
arises at zero temperature.  Locally, such imbalance occurs in the
vicinity of atomic defects. Therefore, in our initial density
matrices, we assume a surplus of spin up (down) density around type-A
(B) vacancies, leading to our lowest energy solution.

Once the self-consistent Hubbard quasi-particle states
$\psi_{n\sigma}(x)$ with energy $E_n$ are obtained, we proceed with the computation of
time-dependent wave functions. Assuming an initial wave packet of the electron injected through one corner of the hexagonal QD (See Fig.1b) as $\Psi(t=0)$ with average energy $\left\langle E_{i} \right\rangle$ of width $\delta E_{i} \sim t_{nn}/2 $, the evolution is given by  $\Psi(t)=\sum
\limits_{n}\left\langle \psi_{n\sigma}|\Psi(t=0) \right\rangle e^{-iE_{n}t/\hbar}\psi_{n\sigma}$. When the time
scale is sufficiently large, $t \gg t_{0}=t_{nn}/\hbar$,
(where $t_{0} \sim$ femto second) the system
reaches a quasi-stationary state from which it is possible to deduce
the localization properties\cite{AdatomAndersonLocaTB1}.

\section{Results and Discussions}

In the following, we first present our tight-binding results which
will serve as a basis to understand the interaction effects within
mean-field approximation discussed in subsection (ii). Finally, in
subsection (iii), we will focus on the interplay between Anderson
localization and magnetization. In this work, we consider defect
concentrations of 1\%, 2\% and 5\%. While the defect sites are chosen
randomly, we consider two idealized cases where the defects are either
perfectly evenly distributed among the sublattices A and B (50-50\%)
or solely distributed in sublattice A (100-0\%). Although these two
cases do not represent realistic situations, they provide a good
physical picture of possible magnetization effects. The tight-binding
calculations are performed for armchair QDs containing 5514, 10806 and
21426 ($\sim$13, 18 and 25 nm QD size, respectively) atoms, while the
mean-field calculations are restricted to 5514 atoms due to computational limits.

\subsection{i) Tight-binding results}

Figure 2 shows the density of states (DOS) in the vicinity of the
Fermi level of a 5514 atoms QD for defect-free and disordered cases
obtained from TB calculations. Black dots represent energies of
interest at which an electron will be injected from the lead. In
particular, as the defect concentration increases, a peak in DOS near
the Fermi level ($E \sim 0.38 eV$) is observed, as
expected. Corresponding time evolution density plots for a $\left\langle E_{i} \right\rangle = 0.38$
eV wave packet are shown in Fig.3, at $t=0$, $t/t_{0}=30$ and
$t/t_{0}=10^6$ (from left to right), for defects concentrations of
2\% (upper panels) and 5\% (lower panels).  Initially, at
$t=0$, we assume that the injected wave packet occupies a
small, defect-free region of the QD. As $t$ is increased, the density
propagates slower for higher defect concentrations, before reaching a
quasi-stationary state above $t/t_{0}=10^4$. At higher time scales,
$t/t_{0}=10^6$ (order of nano second), the wave packet is still
localized around the corner of the QD, especially visible at the
higher defect concentration.

In order to investigate the localization more systematically including
size dependence, in Fig.4 we plot the injected electron's probability
density as a function of distance to the lead corner, integrated over
an angle of $\pi / 3$ (see Fig.1b), and averaged over 20 randomly
generated defect configurations (evenly distributed between sublattice
A (50\%) and B (50\%) for the main frames but unevenly as A (100\%)
and B (0\%) for the inset figure), obtained from TB calculations.
Moreover, time averages over 36 samples between $t/t_{0}=5 \times
10^5$  and  $4 \times 10^6$  were performed.  Here, each column
corresponds to a different size GQD while each row corresponds to a
different defect concentration. Localization lengths denoted by
$\lambda$ were estimated for different injected electron energies (one
near the Fermi level, other two in deep conduction and valence bands),
by logarithmic curve fitting\cite{AdatomAndersonLocaTB1}. At $1\%$ defect
concentration, size effects dominate the densities. Estimated
localization length is larger than the system size even for the
largest QD (25 nm wide) and the energy dependence is weak. As
the defect concentration is increased to $2\%$, we find $\lambda\sim
12$ nm  for 0.38 eV (Fermi level energy) for all QD sizes. At -1.6 and
2.4 eV, $\lambda$ exceeds the system size. Finally, increasing defect
concentration to 5\% decreases localization length to $\lambda\sim
3.5$ nm for $\left\langle E_{i} \right\rangle = 0.38$ eV for all QD sizes. Additionally, we start to
observe localization ($\lambda\sim 6.5$ nm) for the energies $\left\langle E_{i} \right\rangle$ = -1.6 and
2.4 eV. The calculated localization lengths here are consistent with
the TB results by Schubert \textit{et al.}\cite{AdatomAndersonLocaTB1}
obtained for ribbon geometries. Furthermore, as seen in the inset of
Fig.4 , the localization is enhanced for an unbalanced distribution of
100-0\%, consistent with the finding of Ref. \onlinecite{cresti2013broken} obtained for
bulk graphene.

\subsection{ii) Mean-field Hubbard results} 

In previous subsection, we have shown that localization lenght in
armchair QDs can be as low as $\lambda = 3.5$ nm as the disorder
concentration is increased to 5\% for a random but uneven
distribution (50-50\%) among the two sublattices of the honeycomb
lattice, and $\lambda = 2$ nm for the same sublattice distribution. These
localization lengths are lower than the system size of the smallest QD
with 5514 atoms (13 nm wide). In the following, we will focus on 13 nm
QDs in order to investigate the effects of interactions on spin
densities and DOS through self-consistent mean-field Hubbard
calculations.

Figure ~\ref{fig5:result} shows the spin resolved DOS for defects
concentrations of 1\% (upper panels), 2\% (middle panels) and 5\%
(lower panels).  On the left panels, we consider equal number of
randomly distributed defects on A and B sublattices (50-50 \%). Even
though the total spin of such a system is zero as predicted by Lieb's
theorem\cite{LiebsTheorem}, a slight asymmetry can be observed between
spin up and down impurity peaks in the vicinity of Fermi level, due to
broken sublattice symmetry. On the other extreme, if all defects are
placed on sublattice A (right panel), total spin is equal to half of
the total number of defects, and a clear spin splitting is observed in
DOS, a signature of ferromagnetic coupling. As expected, as the
concentration of defects is increased from 1\% to 5\%, impurity peaks
become more pronounced. Again, the black dots represent energy values
of interest which will be used below to calculate the localization
lengths for wave packets with different average energies.

Before discussing the spin-dependent localization properties, in
Fig.~\ref{fig6:result}, we examine spin densities
$n_{i\uparrow}-n_{i\downarrow}$ (upper panels) in terms of defect
positions (lower panels) for different concentrations and sublattice
distributions.  When the system is antiferromagnetic (AFM, for even
number of sublattice A and B defects), statistical distribution of
defects gives rise to formation of magnetic puddles with opposite
signs (shown in red and blue colors online). We note that, a formation
of (non-magnetic) electron-hole puddles due to atomic defects was
previously observed in a TB study of LDOS in large graphene ribbon
structures \cite{AdatomAndersonLocaTB1}. It was found that as the
defect concentration increases from 0.1\% to 1\%, the spatial extent
of electronic puddles is reduced below 1 nm from 5-10 nm. Although the
scale of our magnetic puddle size is consistent with the findings of
Ref.\onlinecite{AdatomAndersonLocaTB1} for 1\% impurity concentration,
we do not observe clear change in puddle size as we increase the
defect concentrations. The formation of magnetic puddles observed in
our calculations is presumably mainly due to the statistical
distribution of defect-induced spins rather than more subtle quantum
interference or interaction effects. We observed similar magnetic
puddle-like structures for other 19 different disorder configurations.
On the other hand, for the uncompensated distribution of defects
(100-0\%), the coupling between the local magnetic moments is
ferromagnetic (FM), as shown in the rightmost panel. Interestingly,
the magnitude of the magnetic moments is almost an order of magnitude
larger than for the FM case. This is somewhat consistent with the
findings of Ref.\onlinecite{AdatomDFTMFH3} where it was found that for defect concentrations larger
than 0.6\%, the antiferromagnetic coupling is suppressed. However, in
our results, we find a reduction of magnetization instead of a
complete suppression, even at high defect concentrations. This difference
is due to the complete randomization of defect positions (for a given
concentration) assumed in our model, unlike the fixed inter-defect
distance model used in Ref.\onlinecite{AdatomDFTMFH3}. We will
investigate this issue further through staggered magnetization
in the next subsection.

 \begin{figure}[t]
\includegraphics[scale=0.25]{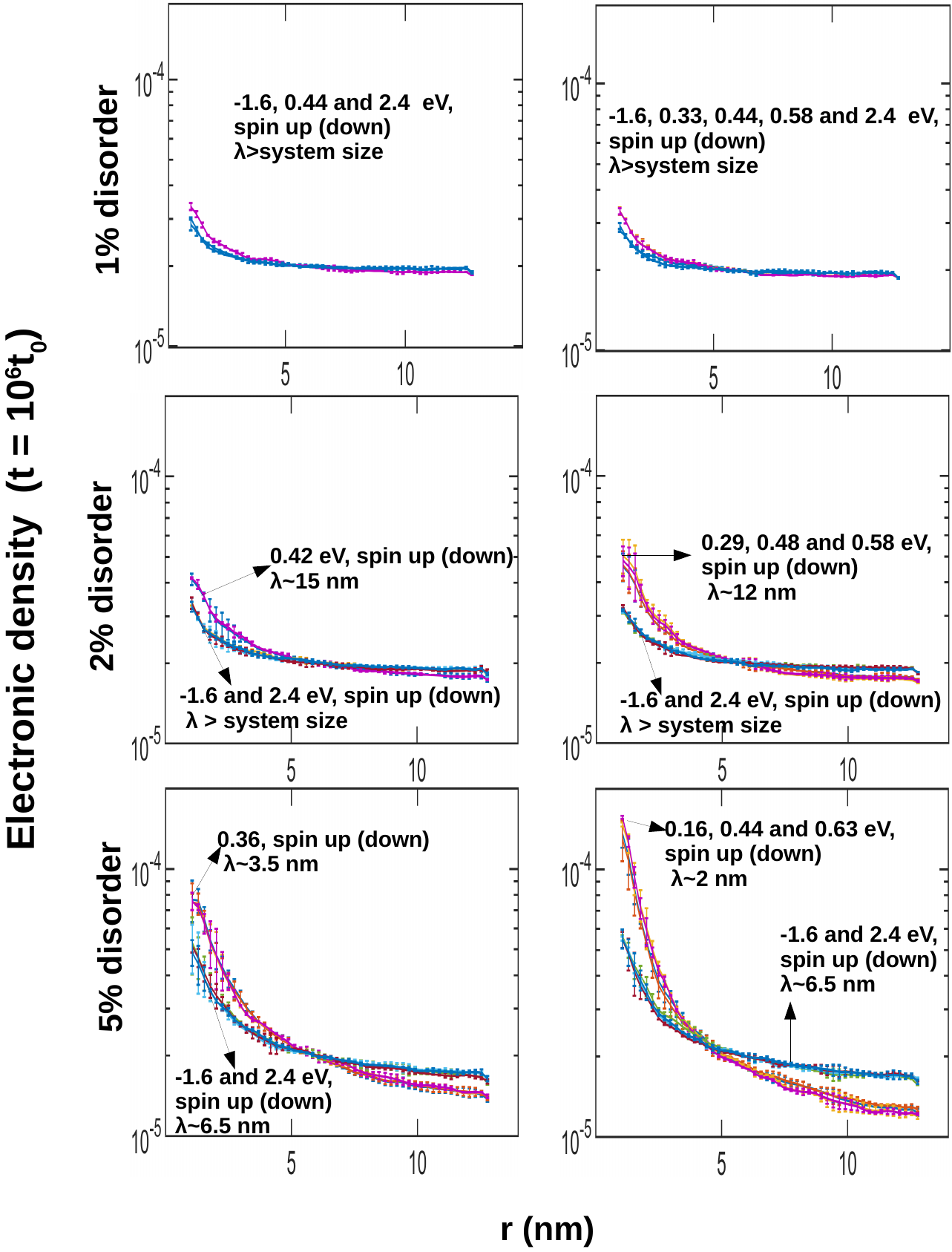}
\caption{(Color online) Time and angle averaged electronic densities
  for different incoming wave packet energies and spins, obtained by
  MFH method. Upper, middle and lower panels correspond to 1\%, 2\%
  and 5\% defects concentrations. Left and right panels correspond to
  AFM (even distribution) and FM (same-sublattice distribution)
  configurations averaged over 20 samples. Spin up and down electrons
  show similar localization behavior. FM configuration shows stronger
  localization except for 1\% disorder where localization lengths are
  larger than the system size.}
  \label{fig7:result}
\end{figure}

\subsection{iii) Localization versus magnetization} 

After having established the extend of disorder induced localization
in within tight-binding model in subsection (i) and the extend of
disorder induced magnetization within mean-field Hubbard model, we now
focus on the relationship between localization and magnetization. In
Fig.7, we plot the angle integrated quasistationary electronic
densities, similar to Fig.4, but obtained using spin-resolved MFH
quasiparticle states. As before, the densities are averaged over 20
configurations and the plots include corresponding error bars. Upper,
middle and lower panels correspond to 1\%, 2\% and 5\% defect
concentrations, respectively, while left and right panels correspond
to evenly (50-50\%) and unevenly (100-0\%) distributed defects among
the two sublattices. Although both spin up and down densities are
plotted in each subfigure, to our surprise no noticeable difference
was found between them, within the statistical error based on 20
randomly distributed configurations. For evenly distributed defects,
the estimated localization lengths from MFH calculations are similar
to those obtained from TB calculations of Fig.4.  Moreover, if the
defects are distributed unevenly among the sublattices, localization
lengths in the vicinity of Fermi level decreases considerably from
$\lambda\sim 15$ nm to $\lambda\sim 12$ nm for 2\% concentration and
from $\lambda\sim 3.5$ nm to $\lambda\sim 2$ nm for 5\% concentration
of defects, consistent with the inset of Fig.4. This is due to the
fact that an even distribution of defects causes more impurity-level
hybridization around the Fermi level compared to uneven distribution
that gives rise to sharper and stronger peak in DOS as seen in
Fig.7. Away from Fermi level, no significant sublattice effect is
observed, as expected.

 \begin{figure}[t]
\includegraphics[scale=0.25]{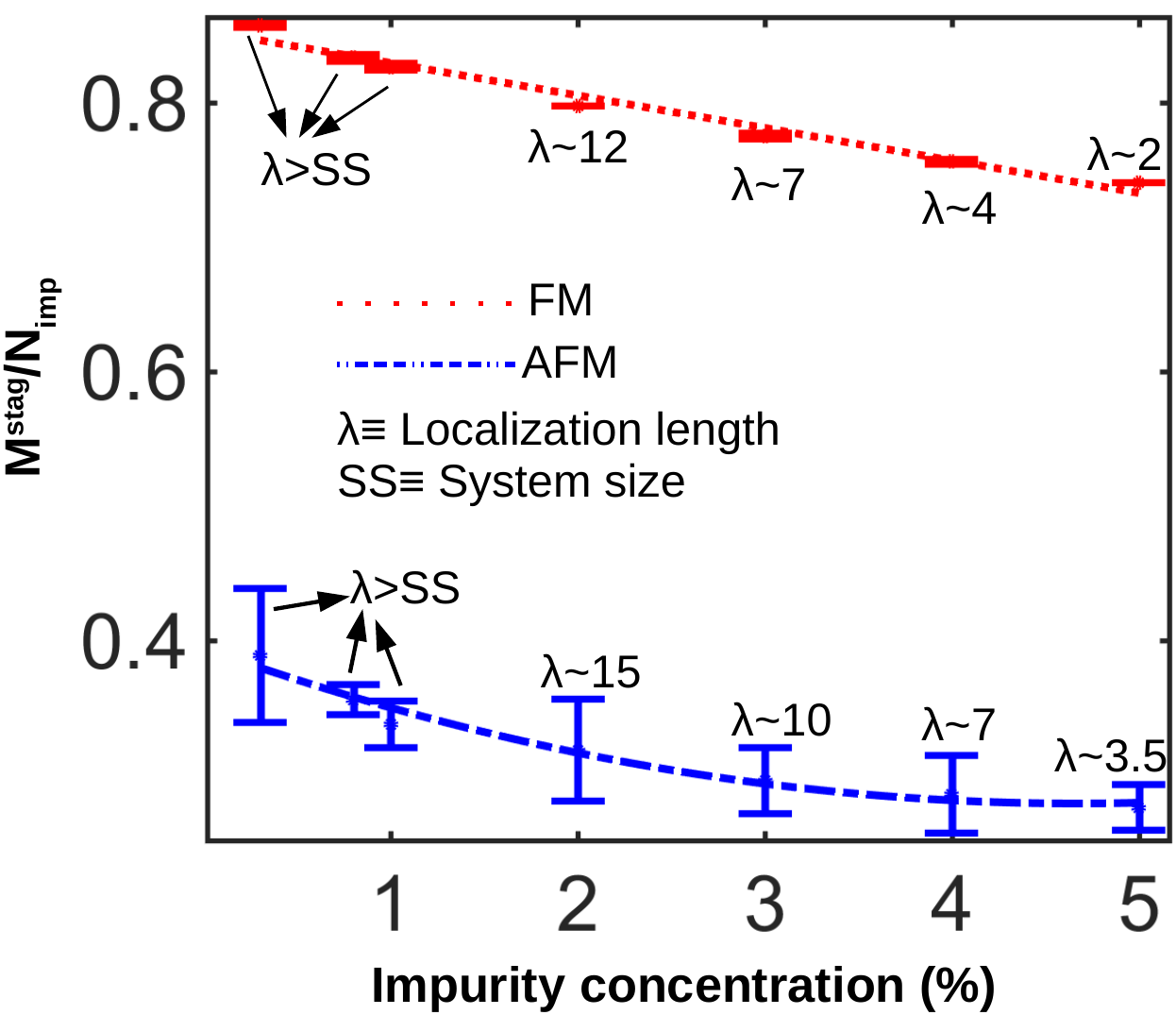}
\caption{(Color online) Staggered magnetization per impurity as a
  function of disorder concentration for AFM and FM configurations.
  The results are averaged over 20 random disorder samples each
  containing 5514 atoms, and corresponding error bars are plotted. For
  data point, corresponding the localization length $\lambda$ is also
  shown if larger than system size  containing 5514 atoms with
  corresponding error bars.}
  \label{fig8:result}
\end{figure}

A useful quantity that describes the magnetic properties is
staggered magnetization defined as :

\begin{equation}
M^{stag}=\sum_i (-1)^x(n_{i\uparrow}-n_{i\downarrow})/2
\end{equation}
where x is even for A and odd for B sublattice sites. In Fig. 8, we
plot the staggered magnetization per impurity, $M^{stag}/N_{imp}$ as a
function of defects concentration of 0.3 - 5 \% for the AFM and FM configurations,
averaged over 20 disordered samples. For each case, the localization
length $\lambda$ is also shown. Several interesting observations can
be made from Fig.8.  First, magnetization of AFM configurations
(same-sublattice defect distribution) is considerably lower than the
FM configurations (even distribution). This reflects the suppression of
antiferromagnetic coupling whenever two impurities are close to each
other \cite{AdatomDFTMFH3}, as discussed above (see Fig.6). Also, the AFM
error bars are much larger than the FM error bars, showing that the
AFM magnetization is more sensitive to the specific distribution of
the defect sites.  Indeed, for some of the samples, large regions
dominated by same-sublattice type defects may be present, causing
weaker AFM suppression. However the net AFM magnetization is never
completely suppressed. Another important observation is that the
localization is consistently stronger for the FM configuration than
for the AFM configurations. This results is consistent with the
conductivity calculations based on tight-binding results of
Ref.\onlinecite{cresti2013broken}, where compensated distribution of defects
in a graphene sheet leads to more localization than the same
sublattice distribution.  Finally, we see that as the defect
concentration increases, the localization length decreases as
expected, and the staggered magnetization per impurity slightly
decreases. Net staggered magnetization of course increases with
increasing number of defects.

\section{Conclusions}

To conclude, we studied random disorder induced localization and
magnetic properties in medium sized hexagonal armchair graphene
quantum dots, using tight-binding and mean-field Hubbard
approaches. We observed magnetic puddle-like formations induced by
random distribution of defects with concentrations between 1\% and
5\%. For QD sizes above 12 nm, defect concentrations of 2\% is needed
in order to observe localization effects. Although the localization
lengths are not directly affected by interactions, we show that, if
the disorder sites are distributed on a same sublattice of the
honeycomb lattice, significantly enhanced magnetism and localization
occurs compared to the evenly distributed antiferromagnetic
case. Surprisingly, no spin dependence of localization length was
observed in neither AFM or FM cases.

\section{ACKNOWLEDGMENT}
This research was supported by the Scientific and Technological
Research Council of Turkey TUBITAK under the 1001 grant project number
116F152, Turkey. The author thanks F. M. Peeters and K. E. \c{C}akmak
for valuable discussions.



\begin{thebibliography}{}
\expandafter\ifx\csname natexlab\endcsname\relax\def\natexlab#1{#1}\fi
\expandafter\ifx\csname bibnamefont\endcsname\relax
  \def\bibnamefont#1{#1}\fi
\expandafter\ifx\csname bibfnamefont\endcsname\relax
  \def\bibfnamefont#1{#1}\fi
\expandafter\ifx\csname citenamefont\endcsname\relax
  \def\citenamefont#1{#1}\fi
\expandafter\ifx\csname url\endcsname\relax
  \def\url#1{\texttt{#1}}\fi
\expandafter\ifx\csname urlprefix\endcsname\relax\def\urlprefix{URL }\fi
\providecommand{\bibinfo}[2]{#2}
\providecommand{\eprint}[2][]{\url{#2}}



%
\bibitem{novoselov2004electric}
 K.S. Novoselov, A.K. Geim, S.V. Morozov,  D. Jiang, Y. Zhang, S.V. Dubonos, I.V. Grigorieva and A.A. Firsov, Science \textbf{306}, 666 (2004).
%
\bibitem{novoselov2005two}
K.S. Novoselov, A.K. Geim, S.V. Morozov, D. Jiang, M.I. Katsnelson, I.V. Grigorieva, S.V. Dubonos and A. A. Firsov, Nature \textbf{438}, 197 (2005).
%
\bibitem{zhang2005experimental}
Y. Zhang, Y. W. Tan, H. L. Stormer and P. Kim, Nature \textbf{438}, 201 (2005).
%
\bibitem{rycerz2007valley}
A. Rycerz, J. Tworzyd{\l}o and C.W.J. Beenakker, Nature \textbf{3}, 172 (2007).
%

\bibitem{Potemski+deHeer+06}
M.~L. Sadowski, G. Martinez, M. Potemski, C. Berger, W.~A. de Heer,
Phys. Rev. Lett. {\bf 97}, 266405 (2006).



%
%
%
%
%



\bibitem{guclu+book14}
A. D. G\"u\c{c}l\"u, P. Potasz, M. Korkusinski, and P. Hawrylak,
Graphene Quantum Dots,
Springer, Berlin, Heidelberg (2014).

\bibitem{Trauzettel+07}
B. Trauzettel, D. V. Bulaev, D. Loss, and G. Burkard,
{Nature} {\bf 3}, 192 (2007).


\bibitem{Schnez+Ensslin+08}
S. Schnez, K. Ensslin, M. Sigrist, and T. Ihn,
Phys. Rev. B {\bf{78}}, 195427 (2008).

\bibitem{Wimmer+10}
M. Wimmer, A. R. Akhmerov, and F. Guinea,
{Phys. Rev. B} {\bf{82}}, 045409 (2010).

\bibitem{Ihn+Ensslin+10}
T. Ihn, J. Güttinger, F. Molitor, S. Schnez, E. Schurtenberger, A. Jacobsen, S. Hellmüller, T. Frey, S. Dröscher, C. Stampfer, K. Ensslin,
{Materials Today} {\bf 44}, 20-27 (2010).

\bibitem{mueller_yan_nanolett2010}
M. L. Mueller, X. Yan, J. A. McGuire, and L. Li,
Nano Lett. {\bf 10,} 2679 (2010).


\bibitem{Hamalainen+Liljeroth+11}
S. K. H\"am\"al\"ainen, Z. Sun, M. P. Boneschanscher, A. Uppstu, M. Ij\"as, A. Harju, D. Vanmaekelbergh, and P. Liljeroth,
{Phys. Rev. Lett.} {\bf 107}, 236803 (2011).


\bibitem{Subramaniam+12}
D. Subramaniam, F. Libisch, Y. Li, C. Pauly, V. Geringer, R. Reiter, T. Mashoff, M. Liebmann, J. Burgd\"orfer, C. Busse, T. Michely, R. Mazzarello, M. Pratzer, and M. Morgenstern,
{Phys. Rev. Lett.} {\bf 108}, 046801 (2012).

\bibitem{Olle+Gambardella+12}
M. Olle, G. Ceballos, D. Serrate, and P. Gambardella,
{Nano Lett.} {\bf 12}, 4431 (2012).

\bibitem{isil+14}
I. Ozfidan, M. Korkusinski,A. D. G\"u\c{c}l\"u, J. A. McGuire and P. Hawrylak,
Phys.Rev.B {\bf 89}, 085310 (2014).



\bibitem{Ezawa+07}
M. Ezawa,
{Phys. Rev. B} {\bf 76}, 245415 (2007).

\bibitem{FRP+07}
J. Fernandez-Rossier and J. J. Palacios,
{Phys. Rev. Lett.} {\bf 99}, 177204 (2007).

\bibitem{Wang+Meng+08}
W. L. Wang, S. Meng, and E. Kaxiras,
{Nano Lett.} {\bf 8}, 241 (2008).
\bibitem{AHM+08}
J. Akola, H. P. Heiskanen, and M. Manninen,
{Phys. Rev. B} {\bf 77}, 193410 (2008).




\bibitem{Guclu+09}
A. D. G\"u\c{c}l\"u, P. Potasz, O. Voznyy, M. Korkusinski, and P. Hawrylak,
{Phys. Rev. Lett.} {\bf 103}, 246805 (2009).


\bibitem{Potasz+10}
P. Potasz, A. D. G\"u\c{c}l\"u, and P. Hawrylak,
{Phys. Rev. B} {\bf 81}, 033403 (2010).


\bibitem{Zarenia+11}
M. Zarenia, A. Chaves, G. A. Farias, and F. M. Peeters,
{Phys. Rev. B} {\bf 84}, 245403 (2011).

\bibitem{ma+12}
W. L. Ma and S. S. Li,
Phys. Rev. B 86, 045449 (2012)


\bibitem{Guclu+Potasz+Hawrylak+2013}
A. D. G\"u\c{c}l\"u, P. Potasz, and P. Hawrylak,
{Phys. Rev. B} {\bf 88}, 155429 (2013).

\bibitem{Szalowski+13}
K. Szalowski, Physica E {\bf 52}, 46 (2013).

\bibitem{altintas+2017}
A. Alt{\i}nta\c{s}, K. E. \c{C}akmak, A.D. G\"u\c{c}l\"u
Phys. Rev. B {\bf 95}, 045431 (2017).

\bibitem{modarresi+guclu17}
M. Modarresi, A.D. G\"u\c{c}l\"u,
Phys. Rev. B {\bf 95}, 235103 (2017).

\bibitem{ozdemir+16}
H. U. Ozdemir, A. Alt{\i}nta\c{s}, A.D. G\"u\c{c}l\"u
Phys. Rev. B {\bf 93}, 014415 (2016).

\bibitem{guclu2016}
A.D. G\"u\c{c}l\"u
Phys. Rev. B {\bf 93}, 045114 (2016).

\bibitem{sevincli+08}
H. Sevincli, M. Topsakal, E. Durgun, S. Ciraci,
Phys. Rev. B, {\bf 77}, 195434 (2008).

\bibitem{guclu+bulut15}
A.D. G\"u\c{c}l\"u, N. Bulut,
Phys. Rev. B {\bf 91}, 125403 (2015).




\bibitem{nanostructures1}
X. Li, X. Wang, L. Zhang, S. Lee and H. Dai, Science \textbf{319}, 1229 (2008).

\bibitem{nanostructures2} 
J. Cai,	P. Ruffieux,	R. Jaafar, M. Bieri,	T. Braun,	S. Blankenburg,	M. Muoth,	A.P. Seitsonen,	M. Saleh,	X. Feng,	 K. M\"{u}llen and R. Fasel, Nature \textbf{466}, 470(2010).

\bibitem{nanostructures3}
M. Treier,	C.A. Pignedoli,	T. Laino,	R. Rieger,	K. M\"{u}llen,	D. Passerone	 and R. Fasel, Nature Chemistry \textbf{3}, 61 (2011).

\bibitem{nanostructures4}
M.L. Mueller, X. Yan, J.A. McGuire and L.S. Li, Nano Letters \textbf{10}, 2679 (2010).

\bibitem{nanostructures5}
Y. Morita,	S. Suzuki,	K. Sato and T. Takui, Nature Chemistry \textbf{3}, 197 (2011).

\bibitem{nanoribbonedge5}
T. Wassmann, A. P. Seitsonen, A. M. Saitta, M. Lazzeri and F. Mauri, Physical Review Letters \textbf{101}, 096402 (2008).



\bibitem{YuanyuanSun+2017}
Y. Sun, Y. Zheng, H. Pan, J. Chen, W. Zhang, L. Fu, K. Zhang, N. Tang and Y. Du, npj Quantum Materials \textbf{2}, 5 (2017).






\bibitem{AdatomExperimet1}
R. Balog, B. J{\o}rgensen, J. Wells, E. L{\ae}gsgaard, P. Hofmann, F. Besenbacher and L. Hornek{\ae}r, Journal of the American Chemical Society \textbf{131}, 8744-8745 (2009).

\bibitem{AdatomExperimet2}
 {\v{Z}}. {\v{S}}ljivan{\v{c}}anin, E. Rauls, L. Hornek{\ae}r, W. Xu, F. Besenbacher and B. Hammer, The Journal of chemical physics \textbf{131}, 084706 (2009).
 
\bibitem{AdatomExperimet3}
 D. C. Elias, R. R. Nair, T. M. Mohiuddin, S. V. Morozov, P. Blake, M. P. Halsall, A. C. Ferrari, D. W. Boukhvalov, M. I. Katsnelson, A. K. Geim and K. S. Novoselov, Science \textbf{323}, 610-613 (2009).

\bibitem{AdatomExperimet4}
J. Balakrishnan, G. K. W. Koon, M. Jaiswal, A. H. Castro Neto and B. {\"O}zyilmaz, Nature Physics \textbf{9}, 284 (2013).

\bibitem{AdatomExperimet5}
H. Gonz{\'a}lez-Herrero, J. M. G{\'o}mez-Rodr{\'\i}guez, P. Mallet, M. Moaied, J. J. Palacios, C. Salgado, M. M. Ugeda, J. Y. Veuillen, F. Yndurain and I. Brihuega, Science \textbf{352}, 6284 (2016).

\bibitem{AdatomExperimet6}
K. M. McCreary, A. G. Swartz, W. Han, J. Fabian and R. K. Kawakami, Physical Review Letters \textbf{109}, 186604 (2012).


\bibitem{Mao+2016}
J. Mao, Y. Jiang, D. Moldovan, G. Li, K. Watanabe, T. Taniguchi, M. R. Masir, F. M. Peeters and E. Y. Andrei, Nature Physics \textbf{129}, 545 (2016).

\bibitem{VacancyExperimet1}
M. M. Ugeda, I. Brihuega, F. Guinea and J. M. G{\'o}mez-Rodr{\'\i}guez, Physical Review Letters \textbf{104}, 096804 (2010).

\bibitem{VacancyExperimet2}
R. R. Nair, M. Sepioni, I. L. Tsai, O. Lehtinen, J. Keinonen, A. J. Krasheninnikov, T. Thomson, A. K. Geim and I. V. Grigorieva, Nature Physics \textbf{8}, 199-202 (2012).

\bibitem{VacancyExperimet3}
Y. Zhang, S. Y. Li, H. Huang, W. T. Li, J. B. Qiao, W. X. Wang, L. J. Yin, K. K. Bai, W. Duan, and L. He, Physical Review Letters \textbf{117}, 166801 (2016).
 
\bibitem{AdatomandvacancyExperimet1}
R. R. Nair, I. L. Tsai, M. Sepioni, O. Lehtinen, J. Keinonen, J. Krasheninnikov, A. H. Castro Neto, M. I. Katsnelson, A. K. Geim and I. V. Grigorieva, Nature Communications \textbf{4}, 2010 (2013).

\bibitem{MetaltoInsulatorExperiment1}
A. Bostwick, J. L. McChesney, K. V. Emtsev, T. Seyller, K. Horn, S. D. Kevan and E. Rotenberg, Physical Review Letters \textbf{103}, 056404 (2009).

\bibitem{AdatomVacancyDFT1}
O. V. Yazyev and L. Helm, Physical Review B \textbf{75}, 125408 (2007).

\bibitem{VacancyMixedTheory1}
N. M. R. Peres, F. Guinea, and A. H. Castro Neto, Physical Review B \textbf{73}, 125411 (2006).

\bibitem{VacancyMFH1}
J. J. Palacios, J. Fern{\'a}ndez-Rossier, L. Brey, Physical Review B \textbf{77}, 195428 (2008).

\bibitem{AdatomMFH1}
B. Uchoa, V. N. Kotov, N. M. R. Peres and A.H. Castro Neto, Physical Review Letters \textbf{101}, 026805 (2008).

\bibitem{AdatomDFT1}
D. W. Boukhvalov, M. I. Katsnelson, A. I. Lichtenstein, Physical Review B \textbf{77}, 035427 (2008).

\bibitem{AdatomDFTMFH1}
D. Soriano, F. Munoz-Rojas, J. Fern{\'a}ndez-Rossier and J. J. Palacios, Physical Review B \textbf{81}, 165409 (2010).

\bibitem{VacancyDFT1}
E. K. Safari, A. A. Shokri, M. BabaeiPour, Journal of Magnetism and Magnetic Materials \textbf{441}, 230-237 (2017).

\bibitem{AdatomDFTTB}
Q. Liang, Y. Song, A. Yang and J. Dong, Journal of Physics: Condensed Matter \textbf{23}, 345502 (2011).

\bibitem{AdatomDFTMFH2}
D. Soriano, N. Leconte, P. Ordej{\'o}n, J. C. Charlier, J. J. Palacios and S. Roche, Physical Review Letters \textbf{107}, 016602 (2011).

\bibitem{AdatomDFTMFH3}
N. Leconte, D. Soriano, S. Roche, P. Ordej{\'o}n, J. C. Charlier and J. J. Palacios, ACS nano \textbf{5}, 3987-3992 (2011).

\bibitem{AdatomAndersonLocaTB1}
G. Schubert and H. Fehske, Physical Review Letters \textbf{108}, 066402 (2012).

\bibitem{LiebsTheorem}
E. H. Lieb, Physical Review Letters \textbf{62}, 1201 (1989).

\bibitem{Anderson}
P. W. Anderson, Physical Review \textbf{109}, 1492 (1958).


\bibitem{hopping1}
S. Reich, J. Maultzsch, C. Thomsen and P. Ordejon, Physical Review B \textbf{66}, 035412 (2002).



\bibitem{dielectricconstant1}
T. Ando, Journal of the Physical Society of Japan \textbf{75}, 074716 (2006).

\bibitem{Slatermatrix1}
P. Potasz, A.D. G\"{u}\c{c}l\"{u} and P. Hawrylak, Phys. Rev. B \textbf{82}, 075425 (2010).

\bibitem{cresti2013broken}
A. Cresti, F. Ortmann, T. Louvet, D. Van Tuan and S. Roche, Physical Review Letters \textbf{110}, 196601 (2013).






\end{thebibliography}
\end{document}